# Diamond Phononic Crystal Spin-Mechanical Resonators with Spectrally Stable Nitrogen Vacancy Centers

Ignas Lekavicius, Thein Oo, and Hailin Wang

Department of Physics, University of Oregon, Eugene, OR 97403, USA

## Abstract

We report the design and fabrication of diamond spin-mechanical resonators embedded in a two-dimensional (2D) phononic crystal square lattice. The rectangular resonator features GHz in-plane compression modes protected by the phononic band gap of the square lattice. A membrane-in-bulk approach is developed for the fabrication of the suspended 2D structure. This approach overcomes the limitations of the existing approaches, which are either incompatible with the necessary high-temperature thermal annealing or unsuitable for 2D structures with the required feature size. Graded soft oxygen etching, with the etching rate decreased gradually to below 1 nm/minute, is used to remove defective surface layers damaged by reactive ion etching. Combining the graded etching with other established surface treatment techniques reduces the optical linewidth of nitrogen vacancy centers in resonators with a thickness below 1 μm to as narrow as 330 MHz.

# I. INTRODUCTION

A spin-mechanical resonator, in which an electron spin couples to a mechanical mode with a high Q-factor, provides an experimental platform for quantum control of both spin and mechanical degrees of freedom and for exploiting mechanical degrees of freedom for quantum information processing. Most of the recent experimental studies on spin-mechanical resonators have employed diamond-based mechanical systems and have used negatively-charged nitrogen vacancy (NV) centers in diamond as a spin system. A variety of mechanical systems such as beams and cantilevers[1-5], bulk acoustic wave (BAW) resonators[6], surface acoustic wave (SAW) resonators[7,8], microdisks[9], and optomechanical crystals[10,11] have been pursued.

NV centers are a promising qubit system for quantum information processing[12-14]. These defect centers feature long decoherence times for electron and nuclear spins, along with high-fidelity optical state preparation and readout[15-19]. The long decoherence time, even at room temperature, dictates that the direct ground-state spin-phonon coupling of a NV center is extremely weak[20,21]. In comparison, the orbital degrees of freedom of the NV excited states can couple strongly to lattice strain induced by mechanical vibrations [20,21]. The NV excited-state strain coupling is about five orders of magnitude stronger than the NV ground-state strain coupling[22-25]. The use of the excited-state strain coupling and hence NV optical transitions, however, also result in a number of technical challenges. First, the resolved-sideband regime, which requires the mechanical frequency to exceed the relevant transition linewidth and is essential for many important spin-mechanical coupling processes[22], is difficult to achieve. For out-of-plane mechanical vibrations used in most spin-mechanical systems, the mechanical frequency (< 20 MHz) is small compared with the linewidth of the NV optical transitions. Secondly, for diamond membranes with a thickness < 1 μm, excessive spectral diffusion of NV centers induced by surface charge fluctuations leads to an effective optical linewidth near or even much greater than 1 GHz at low temperature [26,27], which has been a major technical hurdle for the use of NV centers for spin-mechanical studies as well as cavity QED studies.

In this paper, we report the design and fabrication of diamond spin-mechanical resonators embedded in a two-dimensional (2D) phononic crystal lattice. The resonator features GHz in-plane mechanical compression modes, which are protected by the band gap of the phononic crystal. We have developed a membrane-in-bulk approach, fabricating the suspended 2D phononic structure directly in bulk diamond. This approach overcomes the complexities and difficulties of the existing fabrication approaches[28-34], which are either incompatible with the necessary high-temperature thermal annealing for post-release surface treatment or unsuitable for fabricating 2D structures with the required feature size. To overcome excessive spectral diffusion induced by charge fluctuations near the surface, we have used a graded soft oxygen etching process that gradually reduces the etching rate to significantly below 1 nm/minute. The graded and extremely slow etching process effectively removes defective surface layers damaged by reactive ion etching, without causing significant new damages. We show that diamond surface treatments, incorporating graded soft oxygen etching, can suppress spectral diffusion of NV optical transitions and reduce the optical linewidth of NV centers in resonators with a thickness below 1 μm to as narrow as 330 MHz. These resonators should enable the achievement of the resolved sideband regime for the excited-state strain coupling.

## II. RESONATOR DESGIN

A thin elastic plate is a waveguide for mechanical waves. Since mechanical waves cannot propagate in vacuum, a thin plate with finite dimensions and with free boundaries serves naturally as a mechanical resonator. The eigen modes of the thin plate feature Lamb waves or Lamb modes. We are particularly interested in the in-plane compression modes that are symmetric with respect to the median plane of the plate (i.e. the symmetric compression modes). The mechanical loss of these Lamb modes is in principle limited only by the acoustic attenuation of the elastic material. In practice, the Lamb modes are also subject to anchor loss arising from the tethers that support or attach to the thin plate. However, by embedding the thin plate in a phononic crystal lattice, with the relevant mechanical frequencies in the phononic band gap, we can eliminate the anchor loss of the relevant mechanical modes [35-37].

Figure 1a shows a schematic of the overall design of a thin rectangular diamond plate embedded in a phononic crystal square lattice with a period of 8 μm, along with the displacement pattern of the fundamental compression mode. The frequency of this mode depends primarily on the length and is essentially independent of the thickness of the plate. For the length (9.5 μm) of the diamond resonator used in Fig. 1a, the fundamental mode features a frequency of 0.95 GHz. Figure 1b plots the phononic band structure of the symmetric modes for the square lattice. The large band gap protects and isolates the relevant mechanical modes from the surrounding environment. The numerical calculations for the displacement pattern and the phononic band structure in Fig. 1 were carried out with the use of COMSOL Multiphysics software package. The diamond material parameters used in the calculations include Young's modulus $E = 1050$ GPa, Poisson ratio $\nu=0.2$, and material density $\rho = 3539$ kg/m$^3$.

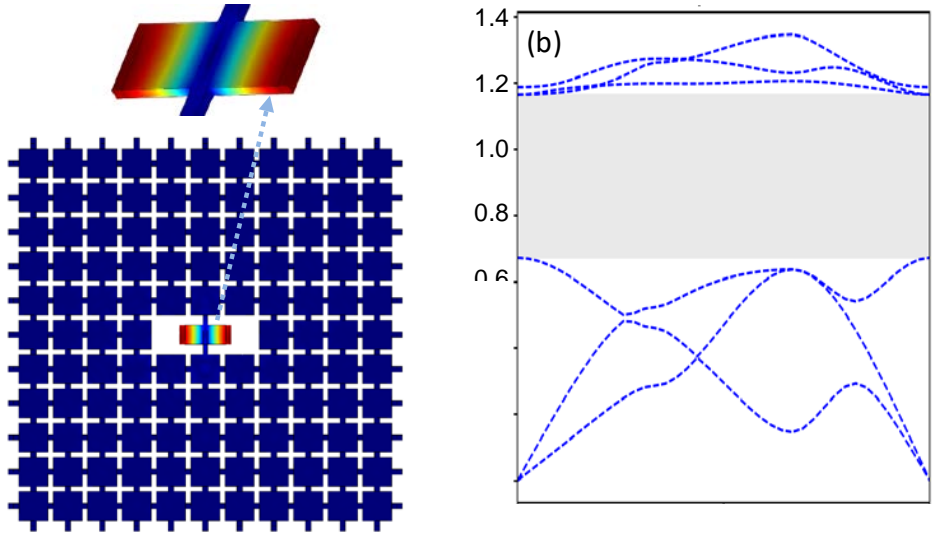

**Fig. 1** (a) The design of a diamond Lamb wave resonator (4.5 μm x 9.5 μm) embedded in a square phononic crystal lattice with a spatial period of 8 μm, along with the calculated mechanical displacement pattern for the fundamental compression mode with a frequency of 0.95 GHz. The dimensions of the bridges in the square lattice are 1.25 μm x 1.25 μm. (b) Phononic band structure of the symmetric modes in the square lattice. The shaded area highlights the phononic band gap.

### III. FABRICATION AND CHARACTERIZATION

As illustrated in Fig. 2a, the fabrication of the 2D diamond phononic structure discussed in Section II starts with surface preparation and ion implantation of a bulk diamond film about 30 μm in thickness for the creation of NV centers near the diamond surface. This is followed by electron-beam lithography (EBL), mask transfer, and reactive ion etching (RIE) of the phononic structure. The diamond film is then thinned down to below 1 μm in thickness with a shadow mask technique (see Fig. 2b), until the phononic structure is completely released. Surface layers that are damaged by the thinning process are then removed through special surface treatment processes. Photoluminescence excitation (PLE) measurements at low temperature are carried out at various stages of the fabrication process for the characterization of the optical coherence or optical linewidth of the NV centers.

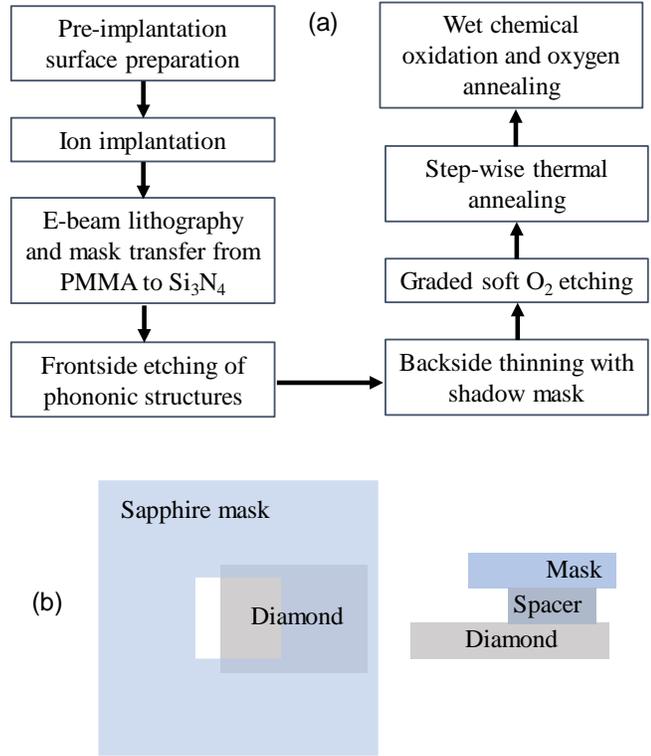

**Fig. 2** (a) A flowchart for the fabrication of phononic crystal mechanical resonators. (b) A sapphire slide with a square hole serves as a shadow mask for backside thinning. A spacer with a thickness of 150 μm between the diamond and the mask is used to avoid trenching.

**A. Creation of NV centers**

Electronic grade chemical-vapor-deposition (CVD) grown single-crystal diamond from Element Six, Inc. was used for the fabrication. The bulk diamond samples were sliced and polished to thin films with dimension (2, 4, 0.03) mm by Applied Diamond, Inc. Inductively-coupled-plasma (ICP) RIE was used for the removal of surface layers damaged by mechanical polishing. A 30 min Ar/$Cl_2$ etch was carried out in a PlasmaPro NGP80 ICP65 etcher from Oxford Instrument, Inc., with an etching rate of 80 nm/minute. This is followed by a 5 min $O_2$ etch in the same etcher and with an etching rate of 100 nm/minute to remove the residual chlorine in the sample. The sample was then cleaned in a triacid solution with a 1:1:1 mixture of sulfuric, nitric, and perchloric acids at 380 °C for about two hours for the removal of surface contaminants. Implantation of $^{15}N^+$ ions at an energy of 85 keV and with a dosage of $3\times10^{10}/cm^2$ was carried out by Innovion, Inc. The kinetic energy used for the implantation leads to a mean nitrogen stopping depth about 100 nm with a straggle about 20 nm[38,39].

After the implantation, we carried out step-wise thermal annealing and additional surface treatments using the procedures developed in an earlier study[39]. The step-wise thermal annealing consists of a 2 hour step at 400 °C, an 8 hour step at 800 °C, and a 2 hour step at 1200 °C, with a temperature ramping rate about 3 °C per minute and with a chamber pressure below or near $10^{-6}$ torr. The thermal annealing was followed by wet chemical oxidation in the triacid solution for 3 hours and annealing in an $O_2$ atmosphere at 465 °C for 2.5 hours. The high temperature thermal annealing leads to the formation of NV centers and repairs to a certain extent damages in the diamond lattice induced by the implantation. The additional surface treatments aim to remove the surface layers graphitized by thermal annealing and to terminate the surface with $O_2$.

Negatively charged NV centers have a permanent electric dipole and are highly sensitive to charge fluctuations in their surrounding environment, which can lead to excessive spectral diffusion of NV optical transition frequencies and can broaden the NV zero-phonon optical linewidth to significantly greater than 1 GHz at low temperature. The zero-phonon optical linewidth or optical coherence of the NV centers can be determined with PLE spectroscopy. For the PLE measurement, we used a tunable diode laser to resonantly excite the NV center through the $E_x$ or $E_y$ transition and a green laser at 532 nm to initialize the NV center to the $m_s=0$ ground spin state and to reverse possible photoionization due to the resonant optical excitation.

Figure 3a shows a single PLE scan of a single NV center obtained at 8 K, after the ion implantation and the extensive surface treatment. Figure 3b shows PLE scans obtained under the same condition, but with repeated scans. The single scan features an optical linewidth near 35 MHz. The additional spectral fluctuations of the NV transition frequency shown in Fig. 3b are in part due to the repumping by the green laser beam. These results are in general agreement with the earlier experimental study[39]. With the surface preparation, ion implantation, and surface treatment procedures discussed above, the NV centers observed consistently feature an optical linewidth near 100 MHz for PLE measurements with repeated scans and with green-laser repumping. However, the experimental results become inconsistent, if $O_2$ plasma etch significantly longer than 5 minutes is used for surface preparation before the ion implantation or if triacid cleaning is not used before the ion implantation. These results indicate that in order to minimize NV spectral diffusion, proper surface preparation before the implantation is as essential as the surface treatment after the implantation.

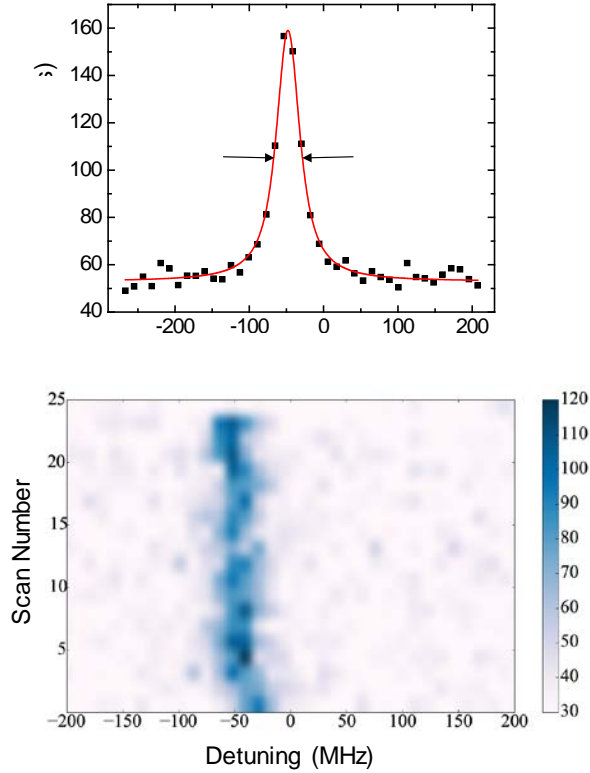

**Fig. 3** (a) A single PLE scan of a single NV center created by ion implantation in a bulk diamond film, followed by surface treatments. The red line is a least square fit to a Lorentzian. (b) Repeated PLE scans of the NV center, with the NV center initialized by a green laser. The data were obtained at 8 K.

**B. Fabrication of suspended 2D phononic structures in bulk diamond**

Three different approaches have been developed for the fabrication of suspended diamond 2D nano-optical and nano-mechanical structures. The diamond-on-insulator approach bonds a diamond film to a silicon wafer[29,31]. 2D diamond structures are released with the wet chemical etching of the $SiO_2$ bonding layer between the diamond and silicon. The membrane-in-frame approach mounts a bulk diamond film in a frame made of materials such as quartz[29]. The membrane can then be etched from the backside (the implanted side is the frontside) until the 2D structure is completely released. For the third approach, angled plasma etching or quasi-isotropic plasma etching is used for undercutting the lithographically defined structures in bulk diamond[32,33]. The undercutting process, however, makes it difficult to fabricate 2D structures with a relatively large feature size (e.g., > 3 μm).

For the fabrication of suspended diamond 2D phononic structures, we have developed an approach that uses a shadow mask during the backside thinning of the bulk diamond film. In this case, a suspended diamond membrane with lithographically defined structures can be fabricated directly in a bulk diamond film, without resorting to undercutting. The membrane is attached directly to the bulk diamond, instead of being supported by a frame made of another material. For this membrane-in-bulk approach, the reuse of unsuccessful samples also becomes possible, as will be discussed in more detail later, which is highly desirable given the relatively poor fabrication yield and the high cost of electronic grade diamond samples.

2D phononic structures were fabricated in a (2, 4, 0.03) mm bulk diamond film. After the cleaning of the diamond film in boiling piranha, plasma-enhanced chemical vapor deposition (PECVD) was used for the deposition of a 280 nm layer of $Si_3N_4$ on the implanted or the front side of the diamond film, on which NV centers have been created about 100 nm beneath the surface. A 10 nm layer of titanium was deposited on the $Si_3N_4$ layer to avoid charging during EBL. A 500 nm layer of Polymethyl methacrylate (PMMA) was then spun onto the sample, followed by EBL that defines the pattern for the 2D phononic structure. Following the photoresist development, $CHF_3$ plasma RIE was used to transfer the pattern to the $Si_3N_4$ layer. The 2D phononic structure pattern was then fabricated onto the diamond film with the $O_2$ ICP-RIE and with the $Si_3N_4$ layer as a hard mask. The etching depth is 1.2 μm, with an estimated etching rate of 100 nm/minute. The etching parameters used include a RF power of 60 W, an ICP power of 420 W, a DC bias of 108 V, a chamber pressure of 10 mTorr, and an $O_2$ flow of 30 sccm. Figure 4a shows an optical image of the 2D phononic structure (the design is shown in Fig. 1a) etched in a bulk diamond film.

The key step in the membrane-in-bulk approach is the use of a U-shaped shadow mask, which is laser-cut from a sapphire slide with a thickness of 150 μm (see Fig. 2b), during the backside thinning of the diamond film. The shadow mask defines the area of the diamond film that will be thinned down and avoids trenching that occurs when the mask is in direct contact with the sample. We positioned the U-shaped sapphire mask with a 150 μm thick spacer at one edge of the diamond film, with an etching area approximately 1.5 mm by 0.75 mm. In this way, the suspended phononic structure is attached or anchored to the bulk diamond film in three sides. If any step of the thinning or post-release surface treatment did not go as expected, the remaining bulk diamond film can still be reused.

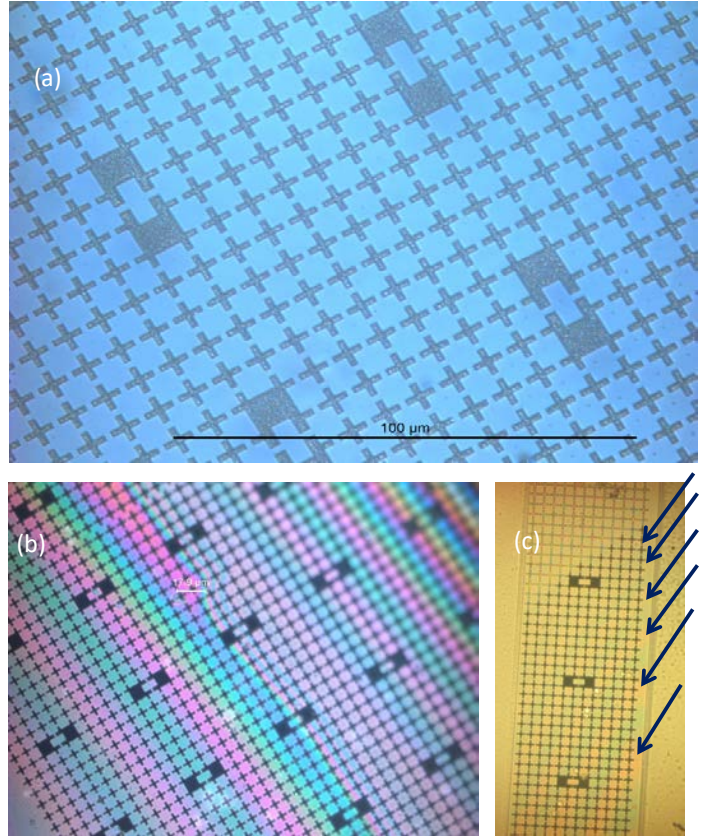

**Fig. 4** (a) Optical image of a 2D phononic structure fabricated in a diamond film with a thickness about 30 μm, for which the structure is not yet released. (b) Optical image of a 2D phononic structure, for which the structure is completely released. (c) Optical image of a 2D phononic structure that contains both released and unreleased regions. The arrows mark the positions of the color fringes. The three images were taken with three different samples.

The diamond film was thinned from the backside with an alternating etch process that consists of 30 minutes of Ar/$Cl_2$ plasma etching, with an etching rate 80 nm/minute, and 2 minutes of $O_2$ plasma etching, with an etching rate 100 nm/minute, followed by 10 minutes of soft $O_2$ plasma etching, with an etching rate 6 nm/minute. The parameters for the Ar/$Cl_2$ plasma etching include a RF power of 210 W, an ICP power of 280 W, a DC bias of 310 V, a chamber pressure of 5 mTorr, 16 sccm $Cl_2$, and 10 sccm Ar. The parameters for the $O_2$ plasma etching are identical to those used for the front side. For the soft $O_2$ plasma etching, the DC bias is set to zero and the RF power is turned off. Other parameters include an ICP power of 500 W, a chamber pressure of 10 mTorr, and an $O_2$ flow of 30 sccm. We ran the thinning process continuously until the phononic structure is released or suspended. This is followed by a long period of soft $O_2$ plasma etching with a gradually decreasing etching rate, as will be discussed in more detail later. We used this graded soft $O_2$ etch to remove the surface layers damaged by the $Ar/Cl_2$ plasma and the hard $O_2$ plasma etching.

Figure 4b shows an optical image of a suspended 2D phononic structure. The bright color fringes are due to slight variations in the thickness of the suspended structure. These variations, which occurred during the slicing and polishing of the bulk diamond film, can be minimized or corrected[29]. The optical interference fringes can be used for the measurement of the membrane thickness[29]. For the 2D phononic structure shown in Fig. 4c, a dividing line between the released and unreleased regions can be clearly identified. This line corresponds to a membrane thickness of 1.2 μm. Nearly parallel color fringes, marked by the arrows in Fig. 4c, can also be discerned. As expected, the fringe visibility increases with decreasing membrane thickness. Each interference fringe or period corresponds to a thickness change of approximately 110 nm. We can thus determine the thickness of a given nanomechanical resonator by counting the number of interference fringes.

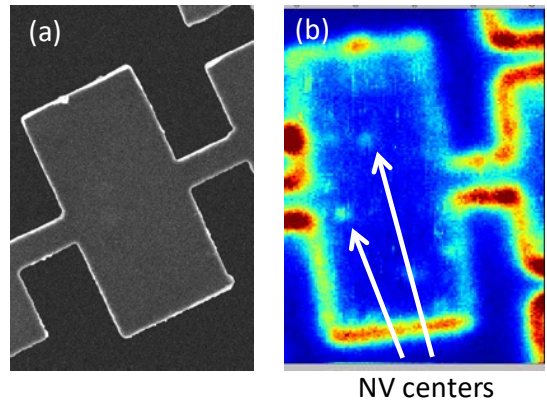

**Fig. 5** (a) A SEM image of a released nanomechanical resonator. The dimension of the resonator is 9.5 μm by 4.5 μm. (b) Confocal optical image showing NV centers in a released nanomechanical resonator.

Figure 5a shows a scanning electron microscope (SEM) image of a released nanomechanical resonator. A confocal optical image of a released nanomechanical resonator showing NV centers in the resonator is also displayed in Fig. 5b. Note that the relatively large-scale edge roughness shown in Fig. 5a is due to an intentional under-etching for the mask transfer from PMMA to $Si_3N_4$, which ensures a relatively thick $Si_3N_4$ layer to protect the NVs from the $O_2$ plasma during the frontside etching. This roughness can be avoided with a more precise mask transfer.

**C. Post-release surface treatment**

A major technical hurdle that has thus far hindered the use of NV centers in spin-mechanical resonators as well as optical microcavities is the severe degradation of optical properties of NV centers in a thin diamond membrane. Previous studies have shown an optical linewidth of 1 GHz or greater at low temperature for NV centers in membranes with a thickness ranging from 1 µm to 100 nm [26,27]. Similar results have also been observed in our samples, which were fabricated with the procedures discussed in Section III.B, but without the graded soft $O_2$ etching step. The samples then underwent the same surface treatments as those used after the ion implantation. For these experiments, fluorescence from individual NV centers in a nanomechanical resonator is detected in a confocal optical setup. These NV centers, which feature permanent electric dipoles, are highly sensitive to charge fluctuations on the back surface of the membrane, even when the back surface is far (e.g., 1 µm) away from the NV centers. We have found no NV centers with a PLE linewidth smaller than 1 GHz when the resonator thickness is less than 1 micron, in agreement with earlier studies. In this case, excessive charge fluctuations in the surface layers damaged by the long etching process lead to strong NV spectral diffusions and thus large optical linewidths. These damages cannot be repaired completely with surface treatment processes such as those used after the ion implantation.

In comparison, for the etching of the 2D phononic structure on the frontside, the $Si_3N_4$ hard mask protects the front surface from damages induced by the $O_2$ plasma. Figure 6a shows a PLE spectrum of a typical NV center in a nanomechanical resonator obtained before the backside etching process. For the average of repeated scans, an optical linewidth of 115 MHz is obtained, which indicates that because of the hard mask, the frontside etching process induces minimal damages on the diamond surface. Note that surface treatments including wet chemical oxidation and oxygen annealing were carried out before the optical characterization.

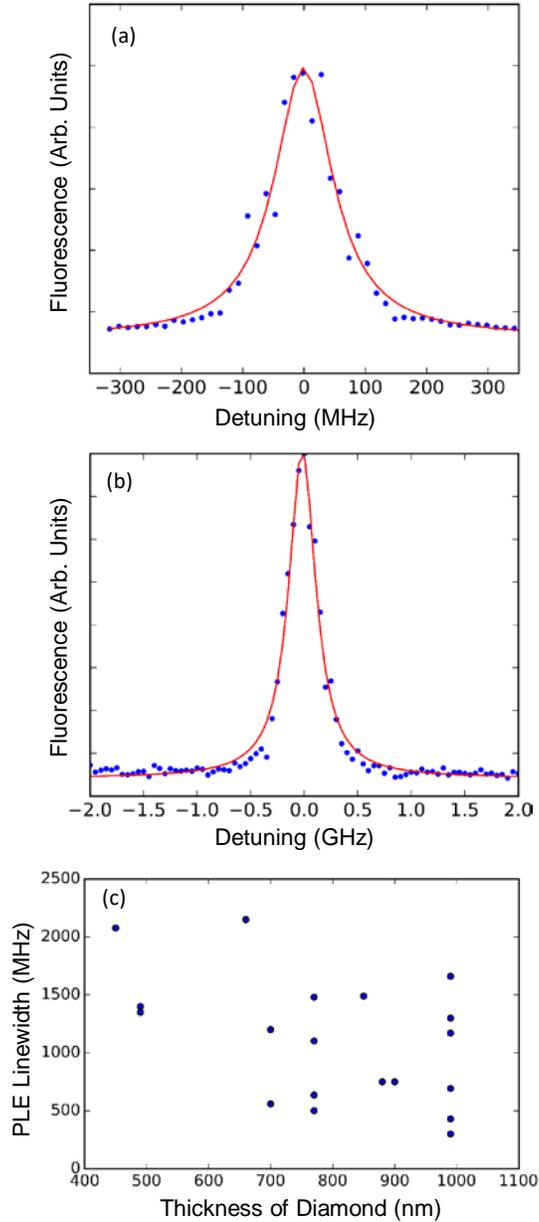

**Fig. 6** (a) PLE spectrum (average of repeated scans) of a NV center in a nanomechanical resonator before the backside etching process. (b) PLE spectrum (average of repeated scans) of a NV center in a released nanomechanical resonator with a thickness slightly below 1 μm. Red lines are least square fits to a Lorentzian, showing a linewidth of 115 MHz for (a) and 330 MHz for (b). (c) A scatter plot of the PLE linewidth for NV centers in the released phononic structures with varying thicknesses, obtained under conditions similar to those in (b). All data were obtained at 10 K.

The technical challenge for the surface treatment after the long reactive ion etching process is to remove the damaged surface layers without causing additional damages. Since wet chemical oxidation is not effective in removing the defective surface layers, we have developed a graded soft $O_2$ etching process, which decreases the etching rate gradually from 6 nm/minute to significantly below 1 nm/minute as the etching progresses. The graded soft $O_2$ etch consists of four steps, a 1 hour etch with an ICP power of 500 W and an etching rate of 6 nm/minute, a 2 hour etch with an ICP power of 200 W and an etching rate of 1 nm/minute, a 10 minute etch with an ICP power of 150 W, and a 10 minute etch with an ICP power of 100 W. This long and graded soft $O_2$ etching process aims to remove the damaged surface layers, while avoiding additional damages through the gradual decrease in the etching rate [40]. The soft $O_2$ etch is followed by the step-wise thermal annealing, triacid wet chemical oxidation, and oxygen annealing, which are essentially the same as those used after the ion implantation discussed in Section III.A [39]. After these extensive surface treatments, we have been able to obtain NV linewidth as small as 330 MHz in a diamond nanomechanical resonator with a thickness below 1 μm, as shown in Fig. 6b. Figure 6c shows a scatter plot of the PLE linewidths (obtained from averages of repeated scans) vs diamond thickness for NV centers measured in released phononic structures. Although a large fraction of the NV centers (>50%) in the released structures still exhibit an optical linewidth near or exceeding 1 GHz, these experiments demonstrate that suitable surface treatments can effectively repair the surface damages induced by the reactive etching process. In principle, the duration and gradual step down of the etching rate for the soft $O_2$ etch can be adjusted and optimized to achieve consistent narrow optical linewidth for NV centers in diamond membranes with a thickness less than 1 μm.

**D. Additional discussions**

We have discussed the optical characterization of NV centers at four different stages of the fabrication process: i) after the implantation, ii) after the frontside etching of the phononic structure, iii) after the backside thinning but before the soft $O_2$ etch, iv) after the soft $O_2$ etch. Surface treatments, including step-wise thermal annealing up to 1200 °C, wet chemical oxidation, and oxygen annealing, have to be carried out before the optical characterization at each of these stages, except that the step-wise thermal annealing is not needed for the optical characterization after the frontside etching of the phononic nanostructures. To increase the yield of and to speed up the fabrication process, we only carry out the surface treatments and thus optical characterization in the last step of a typical fabrication run.

Compared with the three fabrication approaches developed in earlier studies, the membrane-in-bulk approach is straightforward to implement. There are no additional steps of mounting a diamond film to a frame, or bonding a diamond film to a silicon wafer. The membrane-in-bulk approach also allows convenient and direct accesses to both surfaces of the membrane. Furthermore, the $SiO_2$ bonding used in the diamond-on-insulator and the membrane-in-frame approach is not compatible with high-temperature (~ 1200 °C) thermal annealing, which we found is essential for the post-release surface treatment even with the graded soft $O_2$ etch.

A recent experimental study indicated that NV centers with narrow PLE linewidth are formed with the native nitrogen and that NV centers formed from the implanted nitrogen feature much broader PLE linewidth[41]. Additional experimental studies are needed to determine the origin of nitrogen for the NV centers measured in our phononic structures. A conclusive experiment is to measure the optically-detected magnetic resonance (ODMR) response, which identifies whether the nitrogen involved is $^{14}N$ or $^{15}N$, and the PLE response on the same individual NV centers. This can be achieved experimentally with the use of all optical techniques, such as nuclear-spin-dependent coherent population trapping, for the ODMR measurement[42,43].

It should also be noted that out-of-plane mechanical vibrations in commonly used nanomechanical resonators can be conveniently probed with optical interferometric techniques. The interferometric techniques, however, are no longer suitable for high-frequency in-plane mechanical vibrations, such as the compressional modes in a Lamb wave resonator. The in-plane mechanical vibrations can be probed directly through the spin-mechanical coupling, or through near-field optomechanical coupling with whispering gallery mode (WGM) optical resonators. The near-field optomechanical coupling has been used extensively in cavity optomechanics and can be applied to both out-of-plane and in-plane mechanical vibrations[44,45].

A special feature of the phononic crystal mechanical resonator developed in this work is that the resonator can be largely immune to considerable imperfections of the fabricated process as long as the mechanical resonance is in the band gap of the surrounding phononic crystal lattice, because mechanical waves cannot be scattered into vacuum. We do not expect small and smooth variations in the phononic crystal thickness such as those shown in Fig. 4b to affect the mechanical Q-factors. For the 2D square lattice shown in Fig. 1a, the band gap of the phononic crystal is nearly independent of the membrane thickness. Furthermore, the frequency of the fundamental compression mode in our mechanical resonator is determined by the length of the thin plate and is independent of the thickness of the plate, as confirmed by numerical calculations.

## IV. SUMMARY

In conclusion, we have successfully developed and fabricated spin-mechanical resonators that feature a GHz fundamental compression mode and NV centers with optical linewidths as narrow as 330 MHz. The mechanical resonators are embedded in a phononic crystal square lattice with a phononic band gap that protects the relevant compression modes. This type of diamond spin-mechanical resonators has been used in an earlier theoretical study on mechanically mediated spin entanglement and state transfer [46]. The membrane-in-bulk approach developed for the fabrication of the suspended 2D phononic structures, which is relatively easy to implement and allows high-temperature thermal annealing required for post-release surface treatment, can also be extended to the fabrication of diamond 2D photonic structures. A graded soft $O_2$ etching process with the etching rate reduced gradually to significantly below 1 nm/minute has been employed for the removal of surface layers damaged by reactive ion etching without causing significant new damages. Combining the graded soft $O_2$ etch with other established surface treatment techniques can lead to much improved spectral stability of NV optical emissions in diamond membranes with a thickness less than 1 μm. Overall, these advances should enable a new experimental platform for exploiting both spin and mechanical degrees of freedom for quantum information processing, including the use of GHz mechanical compression modes in phononic networks of solid state spins[46].

## ACKNOWLEDGEMENTS

We thank Professor Gaurav Bahl for introducing us to Lamb wave mechanical resonators. This work is supported by AFOSR and by NSF under grants No. 1719396 and No. 1604167.